\documentclass[reprint,prb,aps,amsmath,amssymb,floatfix,superscriptaddress,longbibliography]{revtex4-1}
\usepackage{dcolumn}
\usepackage{bm}
\usepackage{graphicx}
\usepackage{amsthm}
\usepackage{color}
\usepackage{xcolor}
\usepackage{verbatim}
\usepackage{esint}
\usepackage[english]{babel}
\usepackage{amsfonts}
\usepackage{slashed}
\usepackage{latexsym}
\usepackage{bm}
\usepackage[colorlinks = true,
            linkcolor = red,
            urlcolor  = blue,
            citecolor = magenta,
            anchorcolor = red]{hyperref}
\usepackage{esint}
\usepackage{soul}
\usepackage{cancel}
\usepackage[normalem]{ulem}
\usepackage{empheq}
\usepackage{dsfont}
\usepackage{amsmath}
\usepackage{ulem}
\usepackage{epsfig}
\DeclareMathOperator{\sech}{sech}

\begin{document}
\title{ Thermodynamics of Quantum Measurement and the Demon's Arrow of Time}
\author{Kagan Yanik}
\affiliation{Department of Physics and Astronomy, University of Rochester, Rochester, NY 14627, USA}
\author{Bibek Bhandari}
\affiliation{Institute for Quantum Studies, Chapman University, Orange, CA 92866, USA}
\affiliation{Department of Physics and Astronomy, University of Rochester, Rochester, NY 14627, USA}
\author{Sreenath K. Manikandan}
\affiliation{Department of Physics and Astronomy, University of Rochester, Rochester, NY 14627, USA}
\affiliation{Nordita, KTH Royal Institute of Technology and Stockholm University,
Hannes Alf\'vens v{\" a}g 12, SE-106 91, Stockholm, Sweden}
\author{Andrew N. Jordan}
\affiliation{Institute for Quantum Studies, Chapman University, Orange, CA 92866, USA}
\affiliation{Department of Physics and Astronomy, University of Rochester, Rochester, NY 14627, USA}

\date{\today}
	\begin{abstract}
We discuss the thermodynamic aspects of a single qubit based device, powered by weak quantum measurements, and feedback controlled by a quantum Maxwell's demon. We discuss both discrete and time-continuous operation of the measurement based device at finite temperature of the reservoir. In the discrete example where a demon acquires information via discrete weak measurements, we find that the thermodynamic variables including the heat exchanged, extractable work, and the entropy produced are completely determined by an information theoretic measure of the demon's perceived arrow of time. We also discuss a realistic time-continuous operation of the device where the feedback is applied after a sequence of weak measurements. In the time-continuous limit, we derive the exact finite-time statistics of work, heat and entropy changes along individual quantum trajectories of the quantum measurement process, and relate them to the demon's arrow of time.
	\end{abstract}
	\maketitle
	\section{Introduction}
Thermodynamics of quantum measurement powered devices
raises many novel concepts and relations to be explored. 
Although these types of devices and their thermodynamic characteristics have been studied in the past both theoretically \cite{elouard2017extracting,manikandan2019,bhandari2021,janet2016,jordan2019quantum,
elouard2017role,twoqubit2021,manikandan2021efficiently,e22111255,PhysRevLett.122.070603,https://doi.org/10.1002/andp.201700388} and experimentally\cite{PhysRevLett.124.110604,PhysRevLett.125.166802,koski,maillet2019}, there are still plenty of areas that remain untouched. The most important and sought after aspect of these devices are heat exchange, work extraction and the corresponding efficiencies. In this paper, we aim to characterize these quantities 
for a qubit measured via weak quantum measurements and establish
a relation between the thermodynamic quantities, the acquisition of quantum information by a quantum Maxwell's demon, and
the quantum measurement arrow of time~\cite{dressel2017arrow}. To this end, weak quantum measurement will be
revisited in two different operational settings: discrete and time-continuous.

In a closed quantum system where no measurement has been made yet, the dynamics of the system is time reversible. However, when we perform a random weak quantum measurement on the system, we partially collapse the wavefunction and obtain more information about the state of our system. The partial collapse makes the nature of the evolution non-unitary,  and the fact that we have more knowledge about the past state makes it easier for us to distinguish whether the measurement process is more likely to be realized in forward, or in reverse direction, provided the measurement record. Accordingly, the randomness of the measurement process results in a statistically asymmetric inference of the time direction of the evolution of our system \cite{dressel2017arrow,10.1143/PTPS.E65.135,cortes2016spin,PhysRevLett.115.190601}. Distinction between the forward and time reversed evolution can be achieved by a statistical arrow of time which compares the probabilities of the two time directions \cite{dressel2017arrow,jayaseelan2021quantum}.  The relationship between the statistical characteristics of quantum measurement arrow of time and fluctuation relations has been studied both theoretically~\cite{manikandan2019} as well as experimentally in the case of cold atoms~\cite{jayaseelan2021quantum} and superconducting qubits~\cite{harrington2019characterizing}.   Such explorations which are feasible in various qubit-based platforms presently in use further substantiates the timely interest in studying thermodynamic aspects of quantum measurements in terms of the quantum measurement arrow of time. 
\begin{figure}[hbt!]
\includegraphics[width=1\columnwidth]{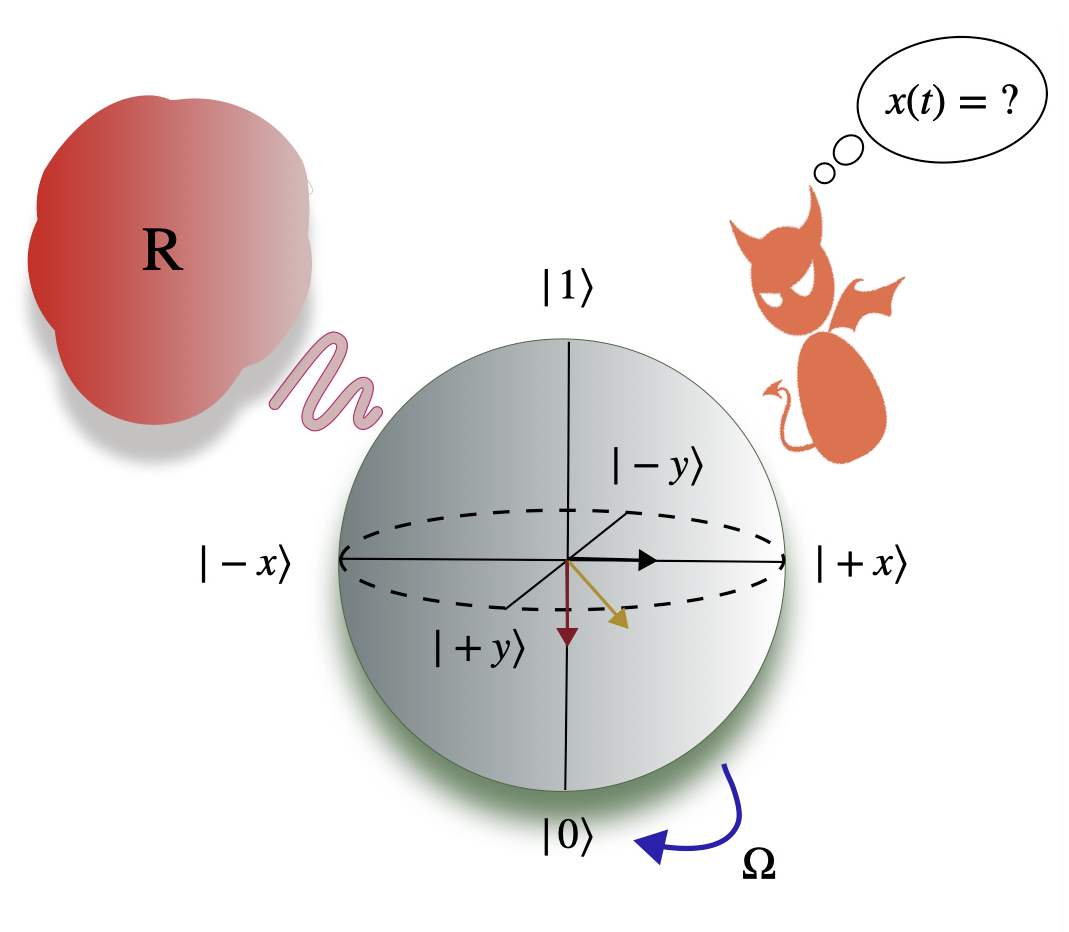}	
 \caption{ The setup for measurement based qubit engine. The qubit is in a thermal state maintained via contact with reservoir ${\rm R}$.   The demon performs an x-measurement increasing the purity and energy of the state. This increase in energy of the qubit can be extracted as work after an optimal feedback (represented by $\Omega$) which brings the qubit back along the negative z-axis.}
\label{fig:setup}
\end{figure}

In this paper, we describe a system that consists of a qubit that is weakly-coupled to a hot thermal
reservoir that thermalizes the qubit consistently to keep it in
a steady state via heat exchange, as demonstrated in Fig.~\ref{fig:setup}. We assume that the hot reservoir 
has a very high heat capacity, with temperature $T$ \cite{blundell_blundell_2006}. For a thermal state, the information about the state 
of the qubit is only in the negative z-axis of the Bloch sphere 
representation. Then we introduce a quantum Maxwell's demon which 
performs a weak quantum measurement on the x-axis of the
Bloch sphere representing the qubit. By performing weak
measurements, the demon acquires new information regarding the 
state of the qubit \cite{Maruyama_2009} 
and the qubit is not in a thermal state anymore.
As a result of the measurement, the Bloch vector gains a new 
vector component along the x-axis and its length changes. This change in length 
of the Bloch vector conjointly changes the energy and purity of 
the qubit \cite{jacobs2003project}. Since an engine is programmed to extract work cyclically from the reservoir, the energy in random form inside the qubit must be 
brought back to the thermal state (negative z-axis) by an external factor. To achieve this, the demon extracts work by an optimal feedback on the system by rotating the qubit around the y-axis of the Bloch sphere at an optimal angular frequency such that the Bloch vector returns back to the negative z-axis~\cite{vijay2012quantum,PhysRevLett.124.110604}.   Subsequent thermalisation brings the qubit to the thermal state.  
 After each measurement, the demon keeps the information in its 
memory, violating the second law of thermodynamics similar to a Szilard engine \cite{plenio2001physics,Maruyama_2009}. However, to
be able to make new measurements cyclically, the demon needs to erase the 
previous information inside its memory via Landauer's erasure 
protocol. This erasure costs the demon, with only a finite memory, a certain amount of work \cite{elouard2017extracting,plenio2001physics,Maruyama_2009,anders2010landauer,PhysRevE.86.040106}.

The process described above for a two-level quantum system has several ingredients that are independently interesting to characterize heat, entropy, and information flows, inviting considerable interest in exploring the fundamental links between these quantities. On the practical side, they also allow us to estimate the finite-time statistics of heat and entropy changes, for example, in terms of the quantum measurement record, which may be directly accessible in an experiment. The present article is aimed at precisely addressing such possibilities, by considering thermodynamic cycles fueled by both discrete, and time-continuous quantum weak measurements. We draw interesting connections between the finite-time statistics of thermodynamic variables such as work, heat, and entropy changes, and relate them to the finite-time statistics of the quantum measurement arrow of time, which can be derived from the quantum measurement record. The Maxwell's demon in this example is the experimentalist making inferences and applying feedback by utilizing the measurement, and computational resources. Therefore, going forward, we may refer to the statistical arrow of time for quantum measurements as the demon's perceived arrow of time, or simply, the demon's arrow of time.

This paper is organized as follows. In Sec.~\ref{sec2}, we characterize a single thermodynamic cycle of the qubit, considering discrete weak quantum measurements. 
We further evaluate the efficiency and coefficient of performance of the device when it acts as a heat engine or a refrigerator respectively. In Sec.~\ref{cont} we discuss the operation of the qubit engine in time-continuous manner.
 We draw our conclusions in Sec.~\ref{conc}.

	\section{Discrete Quantum Weak Measurements}\label{sec2}
We begin by considering thermodynamic cycles on a qubit interacting with a thermal reservoir, fueled by single-step, discrete quantum weak measurements. A similar setup in the strong measurement limit is discussed in Ref.~\onlinecite{elouard2017extracting}. The qubit is initially in a thermal state, $\rho_{i}^{\text{th}}=\exp(-H_{0}/k_{B}T)/Z$, where the free Hamiltonian of the qubit is
$\hat{H}_{0}=\hbar\omega_{0} |1\rangle \langle1|$, and $Z=\text{tr}[\exp(-H_{0}/k_{B}T)]$. The initial energy of the qubit, $E_{0}=\frac{1}{2}\hbar\omega_{0}(1+z_0),$ where $z_0=-1/(2\bar{n}+1)$, and $\bar{n}=1\Big/\left(e^{\frac{\hbar\omega_{0}}{k_{B}T}}-1\right)$ is the thermal occupation of the qubit at temperature $T$.

A Maxwell's demon performs the measurement using an auxiliary qubit that is entangled with our qubit of interest, along its x-axis. An equivalent measurement model is discussed, for example in Ref.~\onlinecite{manikandan2019time}, using a controlled-NOT gate to model the entangling interaction where the auxiliary qubit (the probe) is in a coherent superposition in the computational basis. Given that the probe is also a qubit, the readouts are binary, corresponding to the outcome of a strong measurement in the computational basis of the probe. As a consequence, the model only requires minimum computational resources for the demon to operate, worth a classical bit. The measurement can be described by the two-outcome measurement operators	
$\hat{M}_{+}$ and $\hat{M}_{-}$, which are defined as~\cite{jacobs2003project}:
\begin{equation}
\hat{M}_{\pm}=\frac{1}{2}[(\sqrt{\kappa}+\sqrt{1-\kappa})\ \mathbb{I}\pm(\sqrt{\kappa}-\sqrt{1-\kappa})\hat{\sigma}_{x}],
\end{equation} where $\kappa=1/2-\sqrt{2\gamma^{\prime}\delta t}$ is a dimensionless quantity and an indicator of the strength of the discrete measurement with characteristic measurement rate $\gamma^{\prime}$ and measurement time $\delta t$, which can be related to the resolution of the detector \cite{jacobs2003project}. The measurement operators satisfy the positive operator-valued measure (POVM) relation ${M}^2_{+}+{M}^2_{-}=\mathbb{I}$. These measurements weakly probe the spin state of the qubit along the $x$ direction in the Bloch sphere, discretely. When $\kappa\rightarrow\frac{1}{2}$,  no information is obtained by the demon. In the strong measurement limit, when $\kappa\rightarrow$ 0, 1, maximal discrimination between the eigenvectors ($|+x\rangle$ or $|-x\rangle$) is achieved.

For given $\kappa$, the state of the qubit following measurement outcome
$\hat{M}_{\pm}$ is \cite{wiseman1996quantum,korotkov2011quantum},
\begin{equation}
\rho_{{M}_{\pm}}=\frac{\hat{M}_{\pm}\rho_{i}^{th}\hat{M}_{\pm}^{\dagger}}{P_{f}(\pm)}~~\text{where}~~P_{f}(\pm)={\rm tr}(\hat{M}_{\pm}\rho_{i}^{th}\hat{M}_{\pm}^{\dagger}),
\label{eq:den_after_meas}
\end{equation}
is the forward probability of measurement outcome $\pm$. The statistical irreversiblity of quantum measurements is characterized by the observation that a sequential measurement by the demon can undo the effect of a prior measurement, provided the measurements are time-reversals of each other \cite{korotkov2006undoing,harrington2019characterizing,jordan2010uncollapsing,PhysRevLett.101.200401}. This is accomplished by performing a sequence of $(+,-)$ or $(-,+)$ measurements, restoring the initial state-of-knowledge. The probability of a successful reversal, given the measurement outcome $\pm$ is given by,
\begin{equation}
P_{b}(\pm)=\frac{{\rm tr}(\hat{M}_{\mp}\hat{M}_{\pm}\rho_{i}^{th}\hat{M}_{\pm}^{\dagger}\hat{M}_{\mp}^{\dagger})}{P_{f}(\pm)}. 
\label{eq:backward}
\end{equation} 

The demon's perception of the arrow of time (distinguishability of forward and time reversed measurement) demonstrates the statistical correlation between performing a measurement and undoing it by a sequential measurement \cite{korotkov2006undoing,harrington2019characterizing,jordan2010uncollapsing,PhysRevLett.101.200401}. 
It is defined as the logarithmic ratio of the probability of doing a forward measurement and a time reversed measurement  \cite{dressel2017arrow,harrington2019characterizing},
\begin{equation}
Q(\pm)=\log\bigg(\frac{P_{f}(\pm)}{P_{b}(\pm)}\bigg)=-2\ln(2)-\ln(\kappa(1-\kappa)),
\label{eq:Qj}
\end{equation} 
which, for the example considered here, is independent of the measurement outcome. As $\kappa$$\rightarrow$$\frac{1}{2}$, $Q$$\rightarrow$0, which shows that since no further information is acquired by the demon, the probability of the demon performing a forward weak measurement is same as the demon performing a time reversed weak measurement (it is impossible to distinguish the time direction of the measurement). As $\kappa$$\rightarrow$0 or 1, $Q$$\rightarrow\infty$,  which asserts that the demon acquires maximum possible information in the strong measurement limit.


\subsection{Work extraction}

We now proceed to compute other thermodynamic quantities for the cycle---completed by an optimal feedback and reset via thermalization---in terms of the demon's arrow of time. The average energy of the qubit after the measurement is given by $E_{M}=\frac{1}{2}\hbar\omega_{0}(1+z_0e^{-\frac{Q}{2}}).$ For $\kappa$ $\rightarrow$ 0, 1, maximum information is collected about the x-axis of the Bloch sphere. Hence, the demon generates maximum amount of energy possible, resulting in $E_{M}\rightarrow\frac{1}{2}\hbar\omega_{0}$.
The energy transduced by the measurement process on an average is therefore,
\begin{equation}
Q_{M}=E_{M}-E_{0}=\frac{1}{2}\hbar\omega_{0}z_0\big(e^{-\frac{Q}{2}}-1\big).
\label{eq:delEQ}
\end{equation}
As $\kappa$ $\rightarrow$ $\frac{1}{2}$, no information is collected about the x-axis of the Bloch sphere. Hence, the demon does not give any energy to the qubit, resulting in $Q_{M}\rightarrow0$. 

After measurement, the new length of the Bloch vector is the length of the resultant vector $|z_{f}|=\sqrt{x_{\pm}^{2}+z_{\pm}^{2}}$, where $x_{\pm}$ and $z_{\pm}$ are the coordinates on the Bloch sphere after measurements $\hat{M}_{+}$ or $\hat{M}_{-}$. The magnitude of $z_{f}$ is same for both measurements since both $\hat{M}_{+}$ and $\hat{M}_{-}$ bring an equal change in magnitude on the x-axis, although they have opposite directions; the y-component is still zero after measurement. To extract the most amount of work, the resultant Bloch vector should be rotated around the y-axis with a certain angular (Rabi) frequency $\Omega$ such that it lies entirely on the negative z-axis of the Bloch sphere. This particular rotation around the y-axis is achieved via an optimal feedback \cite{jacobs2003project,vijay2012quantum,PhysRevLett.104.080503,PhysRevA.62.012307}. We assume that the feedback is performed almost instantaneously such that the density matrix after the optimal feedback is given by $\rho_{\text{fb}}=\big(\ I-|z_{f}|\hat{\sigma}_{z}\big)/2$.
\begin{figure}[hbt!]
\includegraphics[width=1\columnwidth]{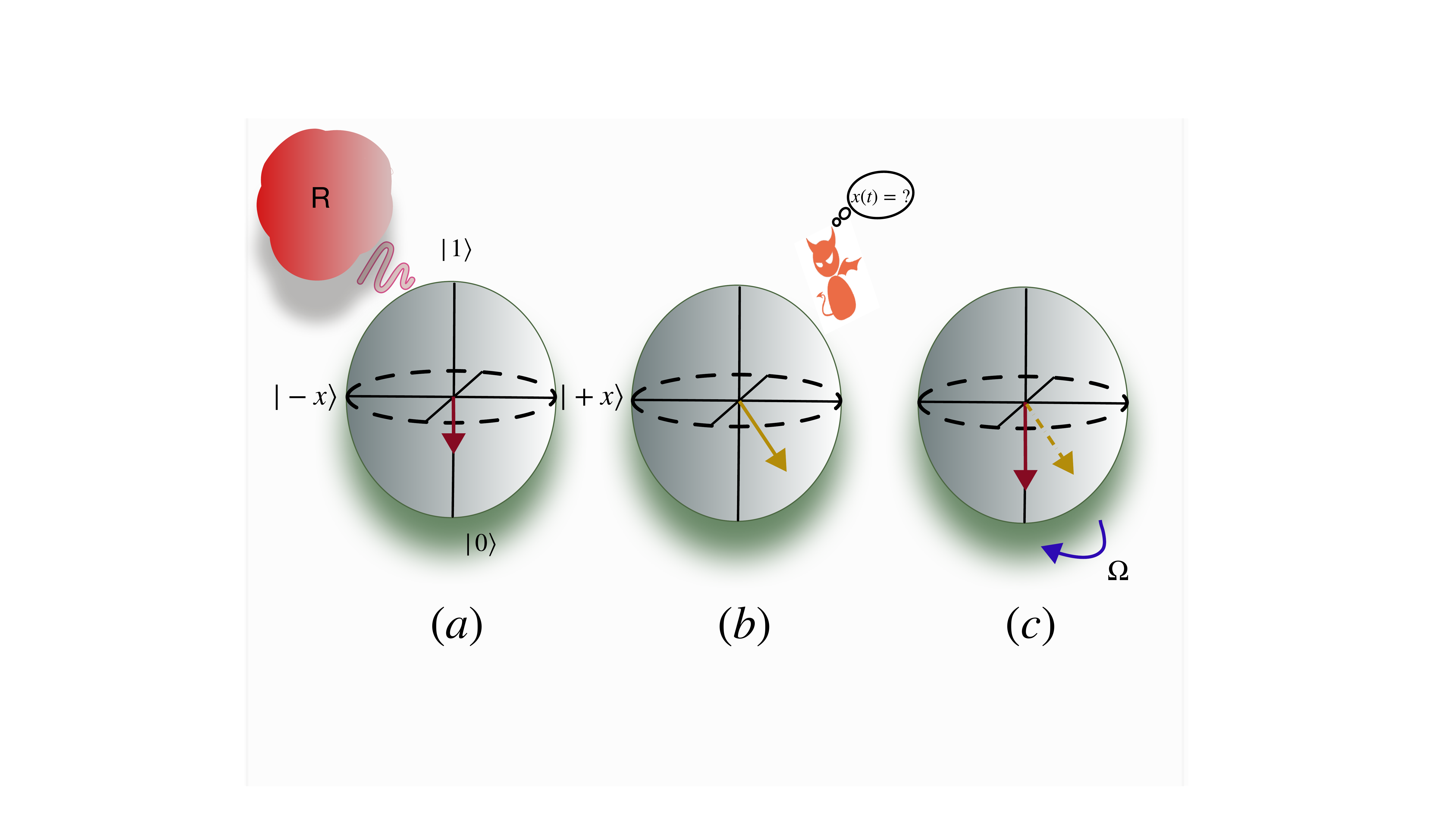}	
 \caption{(a) The hot reservoir thermalizes the qubit at a certain rate via exchange of heat. The red arrow on the qubit is the Bloch vector at the initial thermal state. (b) The demon performs measurement on the qubit changing the length of the Bloch vector (see the orange arrow). The change in length of the Bloch vector also represents the information acquired by the demon. (c) The resultant Bloch vector after measurement (the orange arrow) is rotated by Rabi oscillation characterized by angular frequency $\Omega$. The resultant difference in the length of initial and final Bloch vector is extracted as work.}
\label{fig:sketch}
\end{figure}
The average energy of the system after this feedback is given by 
\begin{equation}
E_{f}=\frac{1}{2}\hbar\omega_{0}\left(1-\sqrt{1+e^{-Q}(z_0^{2}-1)}\right).
\end{equation} Essentially, after the feedback, the Bloch vector is on the negative z-axis and closer to the ground state than the initial state of the qubit. This signifies that the energy of the qubit has decreased, and converted into a form of work that has been extracted by our engine. The average work extracted from measurement after applying optimal feedback is
\begin{equation}
\langle W_{\text{ext}}\rangle=\frac{1}{2}\hbar\omega_{0}\left(z_0e^{-\frac{Q}{2}}+\sqrt{1+e^{-Q}(z_0^{2}-1)}\right).
\label{eq:wextQ}
\end{equation} 
The work extracted is always non-negative regardless of the measurement outcome, and tends to zero when $\kappa\rightarrow 1/2$. In the strong measurement limit, i.e., when $\kappa$$\rightarrow$ 0 or 1, $W_{\text{ext}}$$\rightarrow$ $\frac{1}{2}\hbar\omega_{0}$. 
Such a measurement yields the maximum possible energy transduction and therefore, maximum work extraction.
\begin{figure}[hbt!]
\centering
 \includegraphics[scale=0.45]{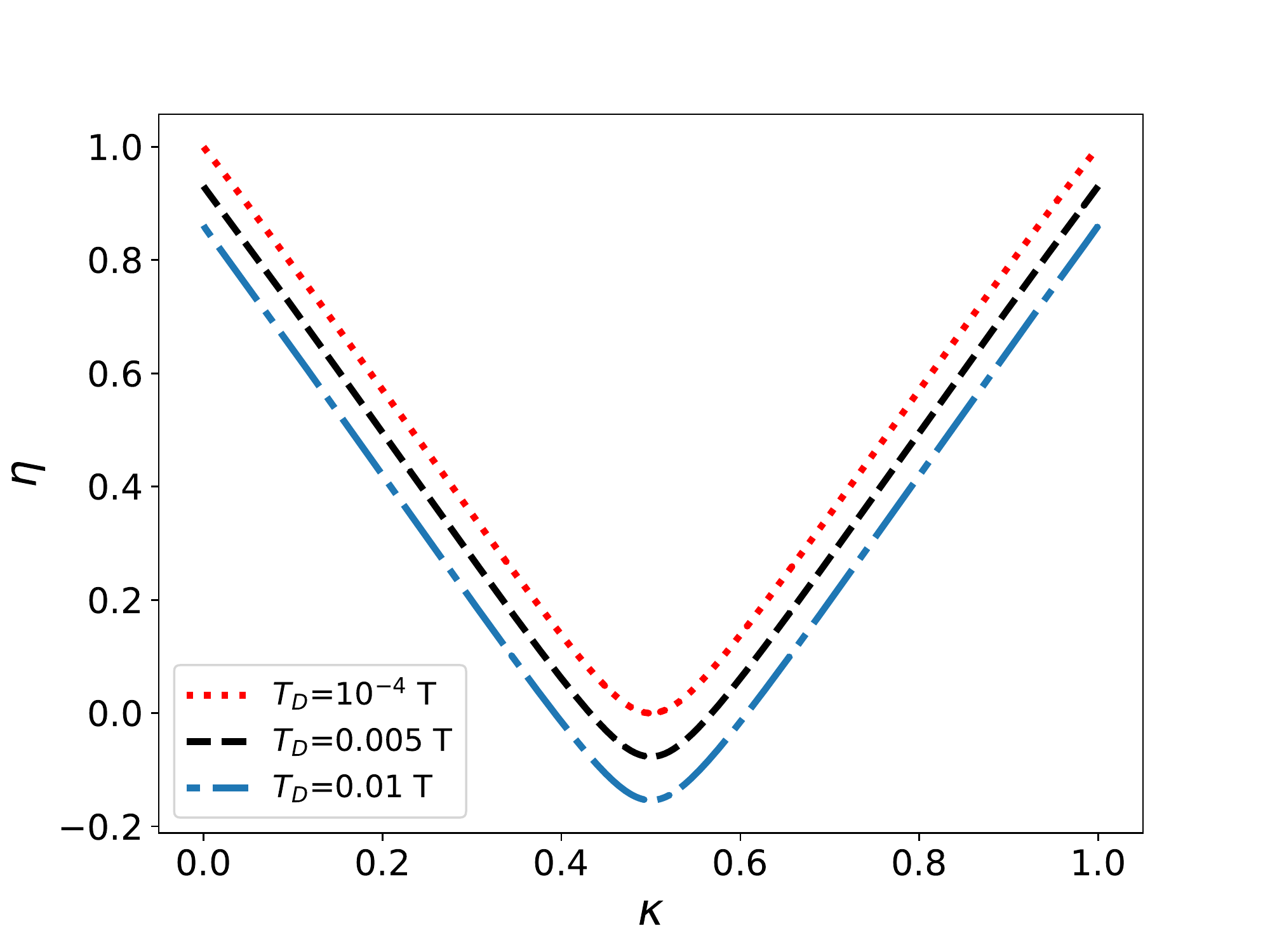}\llap{\raisebox{3.5cm}{\includegraphics[scale=0.18]{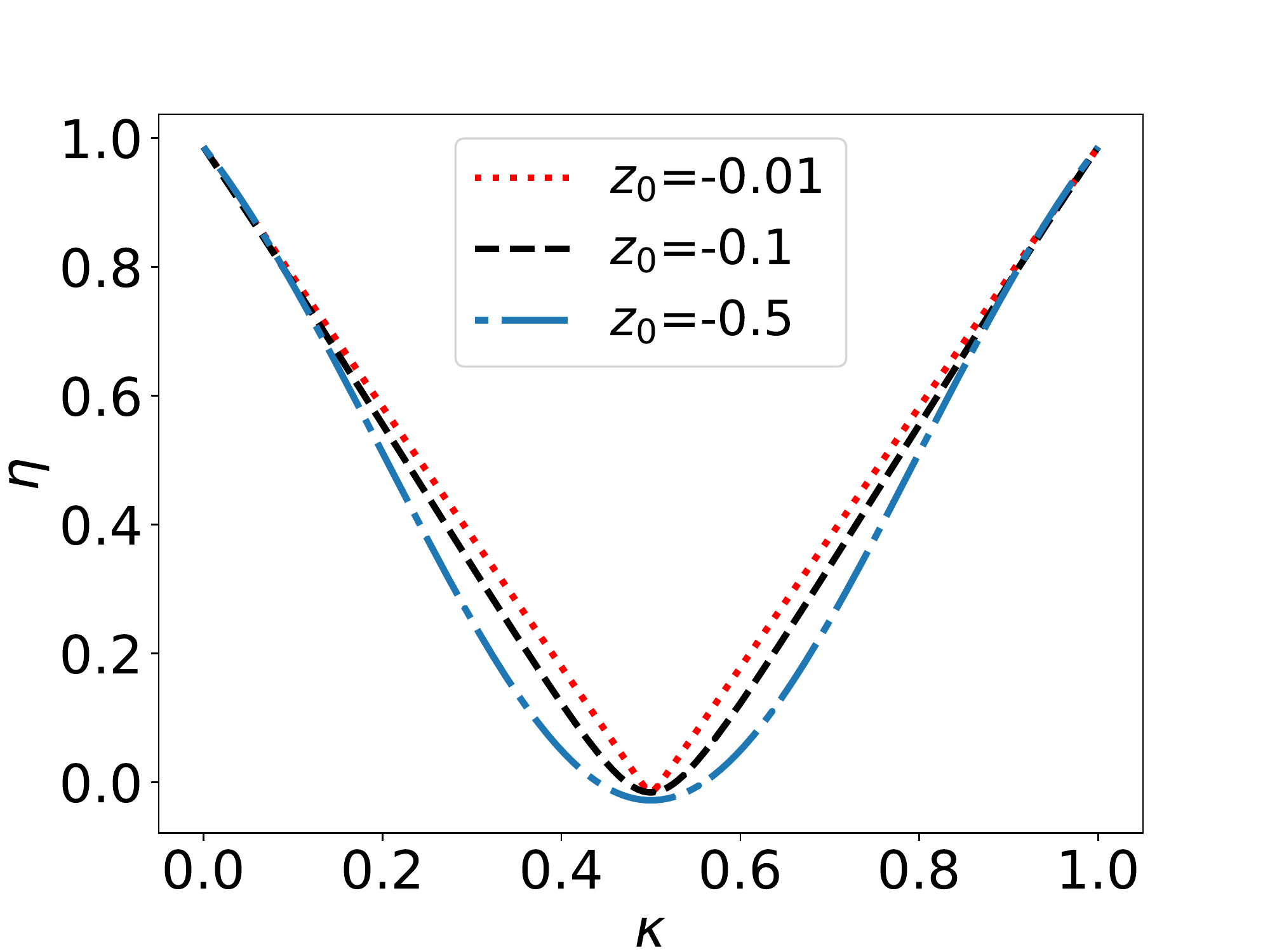}}\hspace{2.7cm}}
 \caption{Efficiency $(\eta)$ of a single discrete measurement as a functon of the measurement strength $(\kappa)$ for three different demon temperatures ($T_D$). For strong measurements ($\kappa\rightarrow0,1$), $\eta\rightarrow1$ whereas for weak measurements ($\kappa\rightarrow\frac{1}{2}$), $\eta\rightarrow 0$. We observe that the efficiency is larger for smaller values of $T_D$. In the inset, we plot the efficiency of a single discrete measurement as a function of the measurement strength for three different initial states. For $0.1\lesssim \kappa \lesssim 0.9$, we observe that the efficiencey increases as $z_0\rightarrow 0$ and takes a maximum value for $\kappa\rightarrow 0,1$ independent of the initial state. For $\kappa$$\rightarrow$$\frac{1}{2}$, the system extracts no work and does not operate as a heat engine anymore. The device acts as a dissipator, yielding negative efficiency. In this plot, we take $\hbar\omega_{0}=0.1 k_{B}T$.}
 \label{fig:effT}
\end{figure}

Since information acquisition of demon violates the second law of thermodynamics \cite{plenio2001physics,Maruyama_2009,anders2010landauer,josefsson2020double}, the memory of the demon (characterized by the measurements) must be erased after each measurement. To formulate the work done to perform this erasure, we follow Landauer's erasure protocol \cite{anders2010landauer,Maruyama_2009,plenio2001physics} and take the number of possible measurements as the number of possible states, resulting in $W_{\text{er}}=k_{B}T_{\mathcal{D}}\log(2)$, where $T_{\mathcal{D}}$ is the temperature of the demon \cite{elouard2017extracting}, satisfying $T_{\mathcal{D}}\ll T$.
Since thermalization happens much slower than weak measurement, the effects of the measurement on the steady state properties of the qubit can be ignored. Thus, the hot reservoir properly thermalizes the qubit only after the feedback is applied.
\subsection{Heat engine and refrigerator}
We define the efficiency of our engine as the ratio between the work extracted after erasure ($W_{\text{ext}}-W_{\text{er}}$) and the heat source ($E_{M}$) \cite{elouard2017extracting,PhysRevA.56.3374}. The efficiency of our Maxwell's demon heat engine can be expressed as
\begin{equation}
\eta=1-\frac{1-\sqrt{1+e^{-Q}(z_0^{2}-1)}+\frac{2}{\hbar\omega_{0}}k_{B}T_{\mathcal{D}}\log(2)}{1+z_0e^{\frac{-Q}{2}}}.
\end{equation}
Two important observations are in order for the qubit measurement engine: (1) The engine can extract non-zero work, even when the reservoir is at zero temperature, by rectifying the measurement induced noise to produce useful work, and (2) The work conversion efficiency $W_{\text{ext}}/E_{M}$ (excluding erasure cost) reaches unity in the strong measurement limit. Both observations suggest quantum advantages in thermodynamic cycles of a qubit, as they result from measurements in a non-commuting basis, as well as feedback rotations through a superposition of states, inaccessible for a classical bit. Similar observations have also been made for a quantum oscillator based measurement engine in Ref.~\onlinecite{manikandan2021efficiently}. As evident from Fig.~\ref{fig:effT}, the measurement engine yields maximum efficiency at maximal measurement strength. For $\kappa\rightarrow 1/2$, the demon obtains no information. Consequently, work extraction tends to zero, the device acts as a dissipator and yields negative efficiency, owing to the erasure cost $W_{\text{er}}$. 
\begin{figure}[hbt!]
\centering
 \includegraphics[scale=0.45]{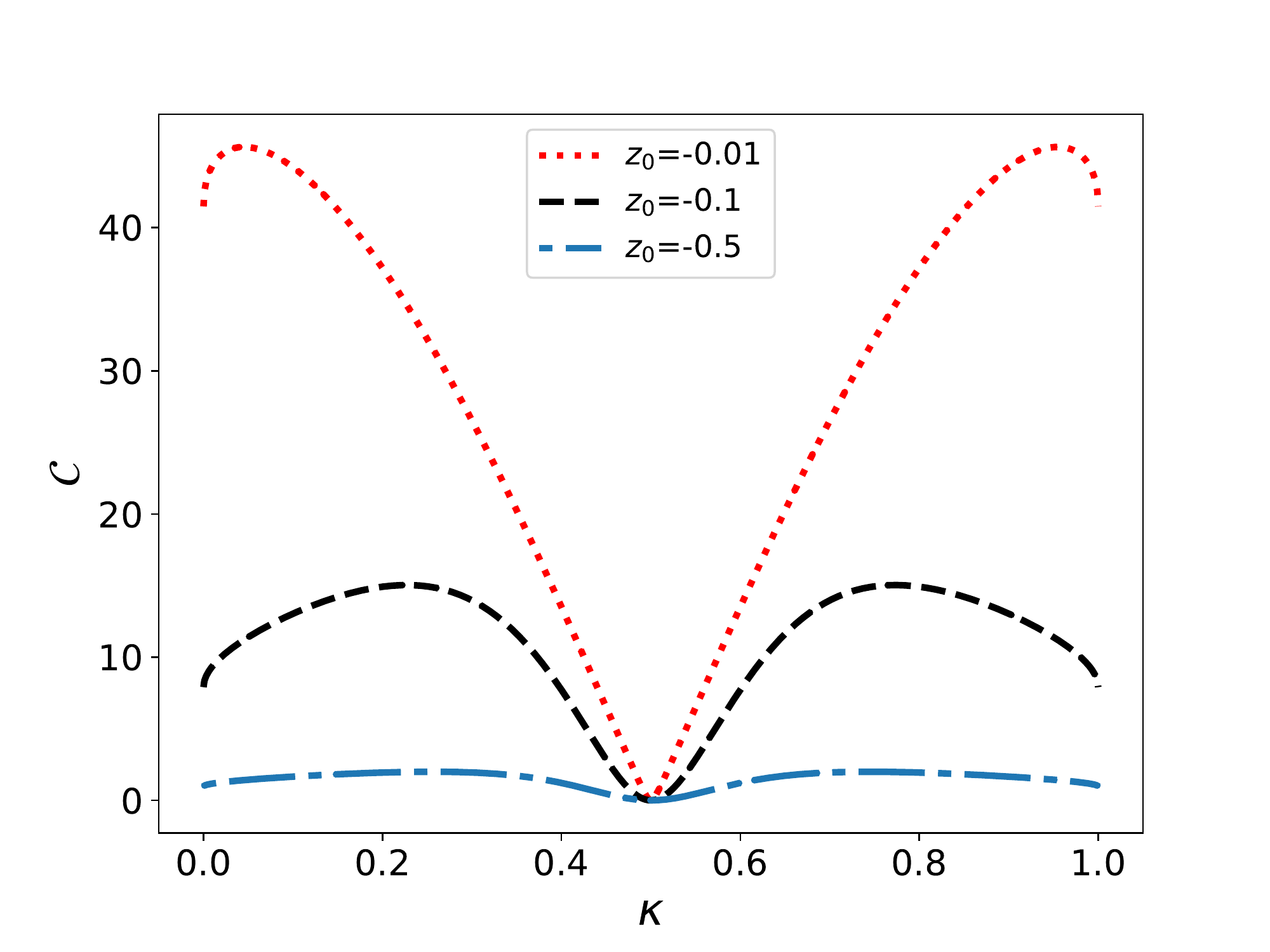}
 \caption{The coefficient of performance for a single discrete measurement varying as a function of the measurement strength for three different initial states ($T_D=0.001 T$ and $\hbar\omega_{0}=0.1 k_{B}T$).}
 \label{fig:COP}
\end{figure}

The thermodynamic cycle above can also be understood as a refrigerator that extracts finite heat from the reservoir. The coefficient of performance ($\cal C$) for the refrigerator is given by
\begin{equation}
    \mathcal{C}=\frac{E_0-E_f}{E_M-E_0+W_{\text{er}}}=\frac{\hbar\omega_0(z_0+\sqrt{1+e^{-Q}(z_0^2-1)})}{\hbar\omega_0 z_0(e^{\frac{Q}{2}}-1)+k_B T_D\log(4)}.
\end{equation}

In Fig.~\ref{fig:COP}, we plot the coefficient of performance ($\cal C$) as a function of $\kappa$ for different initial temperature of the qubit. We observe that the coefficient of performance is symmetric around $\kappa=0.5$ (similar to the case of efficiency, see Fig.~\ref{fig:effT}). However, it is a non-monotonous function of $\kappa$ and shows maximum for a couple of intermediary values of $\kappa$ (placed symmetrically around $\kappa=0.5$) and goes to zero for $\kappa=0.5$.
\begin{figure}[hbt!]
\centering
 \includegraphics[scale=0.45]{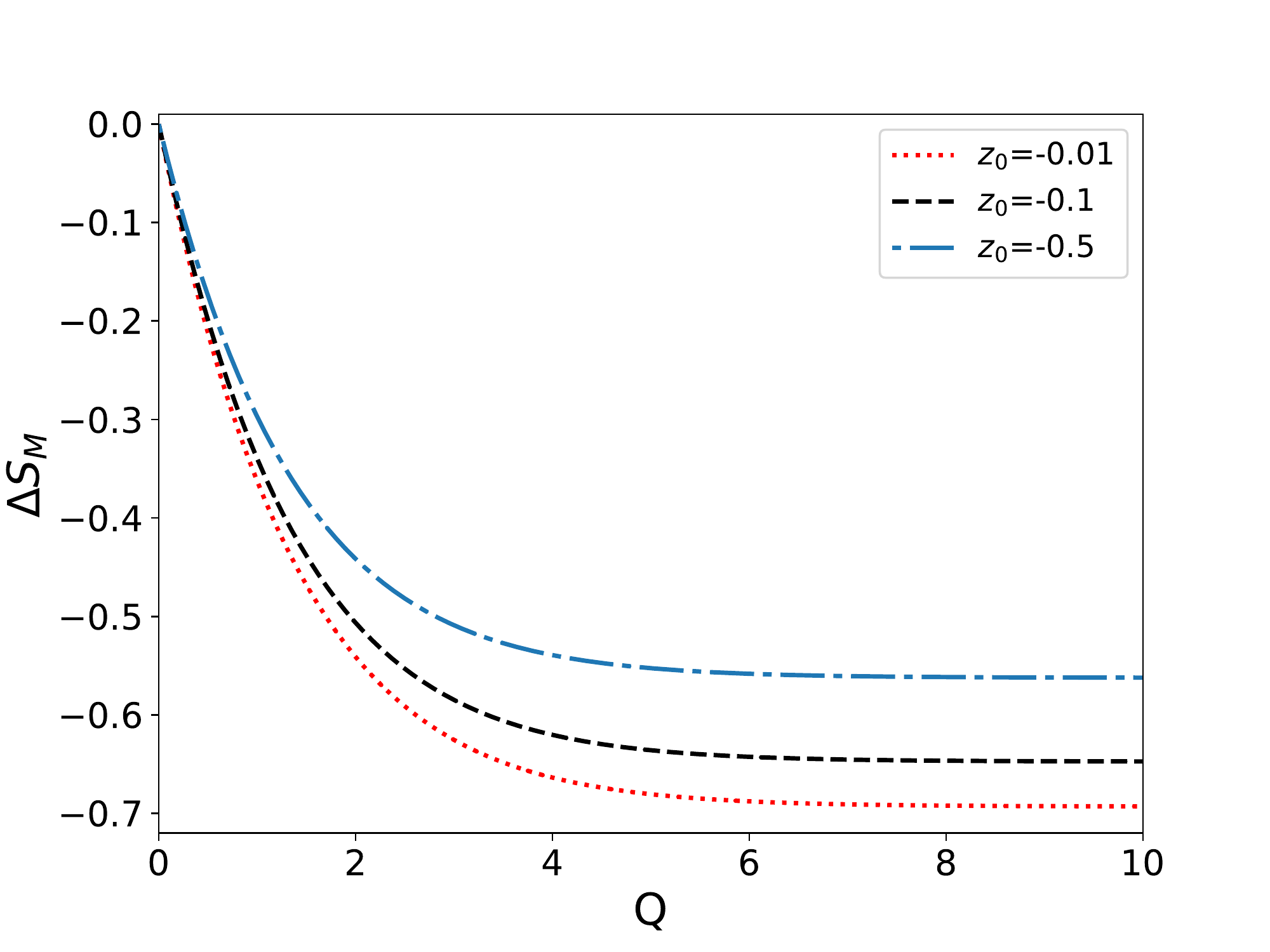}
 \caption{The change in entropy in the process between the measurement and initial state as a function of Q for three different initial states. The parameters are the same as in Fig.~\ref{fig:COP}.}
 \label{fig:dS1}
\end{figure}

\subsection{ Entropy changes}

Here we explore the net entropy changes for the qubit in a cycle. The change in entropy in the measurement process, from preparation to measurement would be given by $\Delta S_{M}=S[\rho_{M\pm}]-S[\rho_i^{\text{th}}]:$
\begin{equation}
\Delta S_M=\frac{1}{2}\left(Q+\gamma (0)-|z_f| \ln\left[\frac{1+|z_f|}{1-|z_f|}\right]\right),
\label{dS1}
\end{equation}
where $\gamma(0)=z_0\ln({(1+z_0)}/{(1-z_0)})$ depends on the initial temperature of the qubit via $z_0$, and $|z_{f}|=\sqrt{1+4\kappa(1-\kappa)(z_0^{2}-1)}$ is the length of the resultant Bloch vector following measurement. The information kept by the demon changes in the process of erasure and hence changing the associated entropy, $\Delta S_{\text{er}}=k_{B}\log(2)$. Eq.~(\ref{dS1}) shows that the change in entropy depends on two components: the $Q$ term depends on  the trajectory of the qubit unique to the measurement and $z_0$ and $z_f$ terms are boundary contributions. As shown in Fig.~\ref{fig:dS1}, the change in entropy production associated with measurement ($\Delta S_M$) is a monotonously decreasing function of $Q$, bounded from above by $Q/2$. Note that, unitary rotation associated with the feedback process generates no entropy production.

\begin{figure}[hbt!]
\centering
 \includegraphics[scale=0.45]{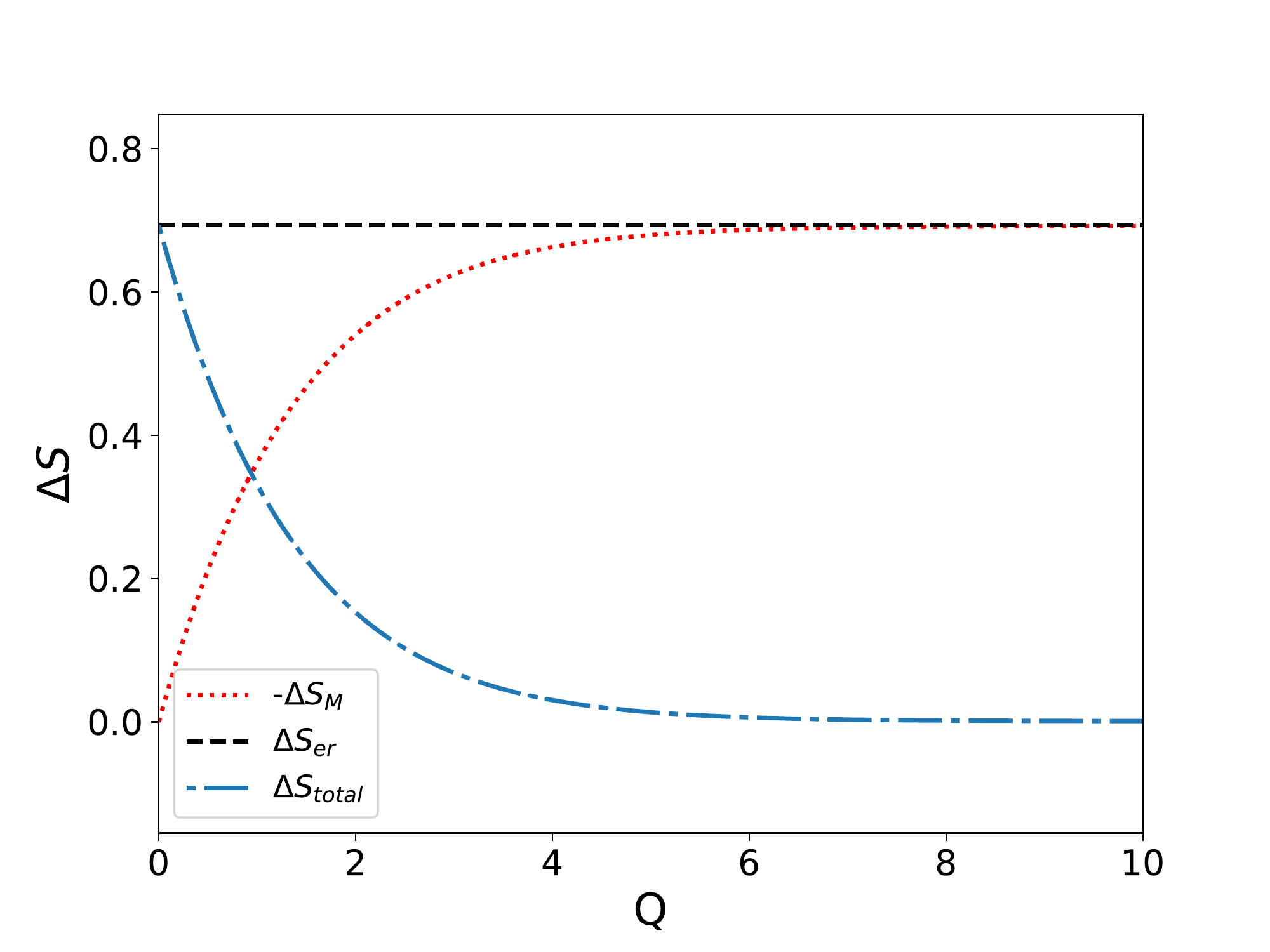}\llap{\raisebox{2cm}{\includegraphics[scale=0.2]{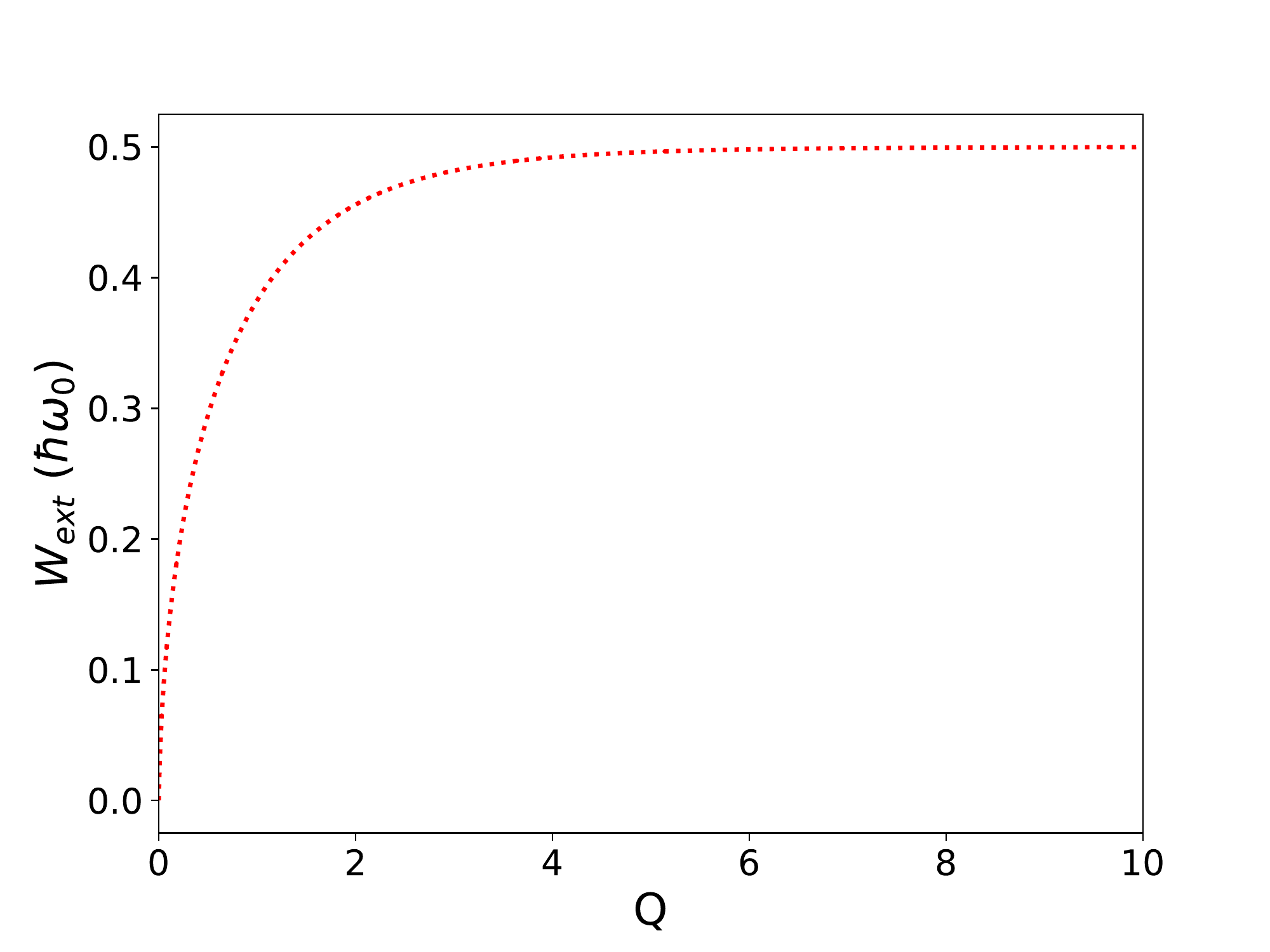}}\hspace{1cm}}
 \caption{The change in entropy in the process between the measurement and initial state (red, dotted line), and the total change in entropy (blue, dashed line) and the change in entropy associated with the process of erasure (black, dashed line) as a function of Q for $z_0=-0.05$. In the inset, we plot the extracted work extracted as a function of Q for $z_0=-0.05$. Other parameters are the same as in Fig.~\ref{fig:COP}.}
 \label{fig:dS}
\end{figure}

In Fig.~\ref{fig:dS}, we observe that for $Q\gtrsim7$, $\Delta S_M$ exactly cancels out $\Delta S_\text{er}$ giving zero net entropy production. An optimal quantum measurement based engine (with least possible dissipation) is achieved in a regime, where the measurement can extract  the maximum amount of work (see the inset of Fig.~\ref{fig:dS}). When the measurement produces no extraction, all the work done goes to dissipation.

\section{Continuous Weak Measurements}\label{cont}
We now proceed to discuss time-continuous version of the measurement engine for which the demon performs a sequence of weak quantum measurements prior to applying the feedback.  As in the discrete case, the qubit is initially attached to a thermal reservoir, but for the probe, we consider a continuous variable system, for example a photon undergoing collisional interactions with a superconducting qubit, whose quadrature is subsequently measured (homodyne measurement). Such time-continuous quantum measurements have been studied extensively in literature using different theoretical tools\cite{chantasri2015stochastic,chantasri2018simultaneous,lewalle2017prediction,PRXQuantum.3.010327,PhysRevA.36.5543,manikandan2019time}, and experimental demonstrations have been achieved\cite{weber2014mapping,vijay2012quantum,harrington2019characterizing}.  An ensemble of identically prepared photons may arrive sequentially, scatter off the qubit and get homodyne-detected, implementing a sequence of weak quantum  measurements. The time-delay between passage of photons ($\delta t$) can be small enough (within the resolution of the detector) such that a realistic time-continuous limit exists. If the measurements were to continue for a duration much longer than the characteristic measurement time ($\tau$), the qubit collapses to one of the eigenstates of the measured observable. We assume that the measurements will be performed in a time-scale much faster than the thermalization time. The work extraction is similar to as before. After a sequence of continuous weak measurements, a feedback rotation is applied for extracting work. In addition to making connections to a well-studied time-continuous limit of weak quantum measurements for the engine's thermodynamics, such an analysis is also timely given the feasibility of implementing real-time quantum feedback (work extraction), for example, in the superconducting platform~\cite{vijay2012quantum}. Additionally, the model also serves to describe both cold atom~\cite{jayaseelan2021quantum}, and superconducting platforms~\cite{harrington2019characterizing}, where the fluctuation relations for the quantum measurement arrow of time have been probed in experiments.

The time-continuous weak quantum measurements of $\hat{\sigma}_{x}$ for the forward and backward measurements are described by the Kraus operators \cite{PhysRevLett.126.100403,dressel2017arrow},
\begin{align}
&\hat{M}_{F/B}=\bigg(\frac{\delta t}{2\pi\tau}\bigg)^{\frac{1}{4}}e^{-\frac{\delta t(r\mp\hat{\sigma}_{x})^{2}}{4\tau}},
\label{Mcont}
\end{align}
respectively. The backward Kraus operator ($M_B$) comes from the measurement result $r_B = - r_F$, corresponding to ``inverting" measurement outcome that would erase the information in the forwards measurement~\cite{dressel2017arrow}. Equivalently, this corresponds to measuring the time-reversed operator $\Theta \sigma_x \Theta^{-1} = - \sigma_x$, where $\Theta$ is the time-reversal operator.  Here $\delta t$ is the time spent between measuring two readouts while $\tau$ is the characteristic measurement time taken to separate the two
Gaussian distributions by two standard deviations \cite{chantasri2015stochastic}. 
The measurement yields a normalized readout value $r$, which in simulations is sampled from two Gaussian distributions with mean values $+1$ (pointing towards the $|+x\rangle$ eigenvector) and $-1$ (pointing towards the $|-x\rangle$ eigenvector) and variance ${\sqrt\frac{\tau}{dt}}$. Given this we also expect fluctuations in work extraction and efficiency. Thermodynamic cycles can be constructed similar to the discrete quantum weak measurement example we discussed before, and our objective again is to explore connections between thermodynamic and information theoretic variables of interest. 

Recall that, in the discrete example, the work, heat and entropy changes did not have fluctuations, and therefore their statistics were straightforward. A crucial difference in the time-continuous limit is that the statistics of work, heat, and entropy changes are not the same for individual realizations of the measurement process. Computing their probability distributions corresponds to deriving exact finite-time statistics of thermodynamic variables, which has gained lots of interest in recent years in the stochastic thermodynamics of nanoscale classical systems~\cite{schmiedl2007optimal,seifert2012stochastic,manikandan2018exact}.
\subsection{Finite-time statistics of work, heat, and entropy changes}
We now proceed to derive the exact finite-time statics of work, heat and entropy changes, given that time-continuous measurements of interest in this section are fundamentally stochastic quantum processes of finite duration.
\begin{figure}[!htb]
\centering
 \includegraphics[scale=0.45]{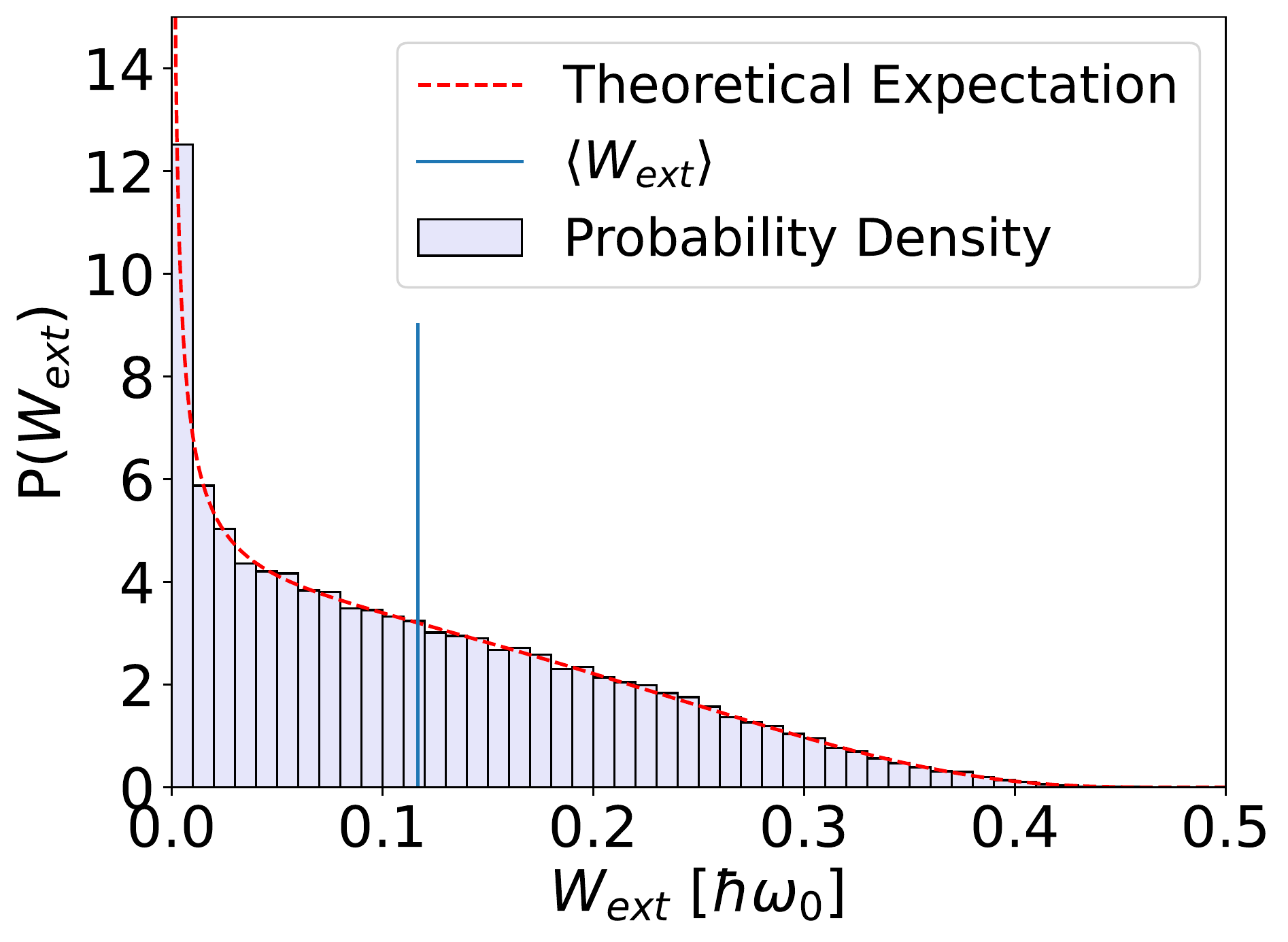}
  \includegraphics[scale=0.45]{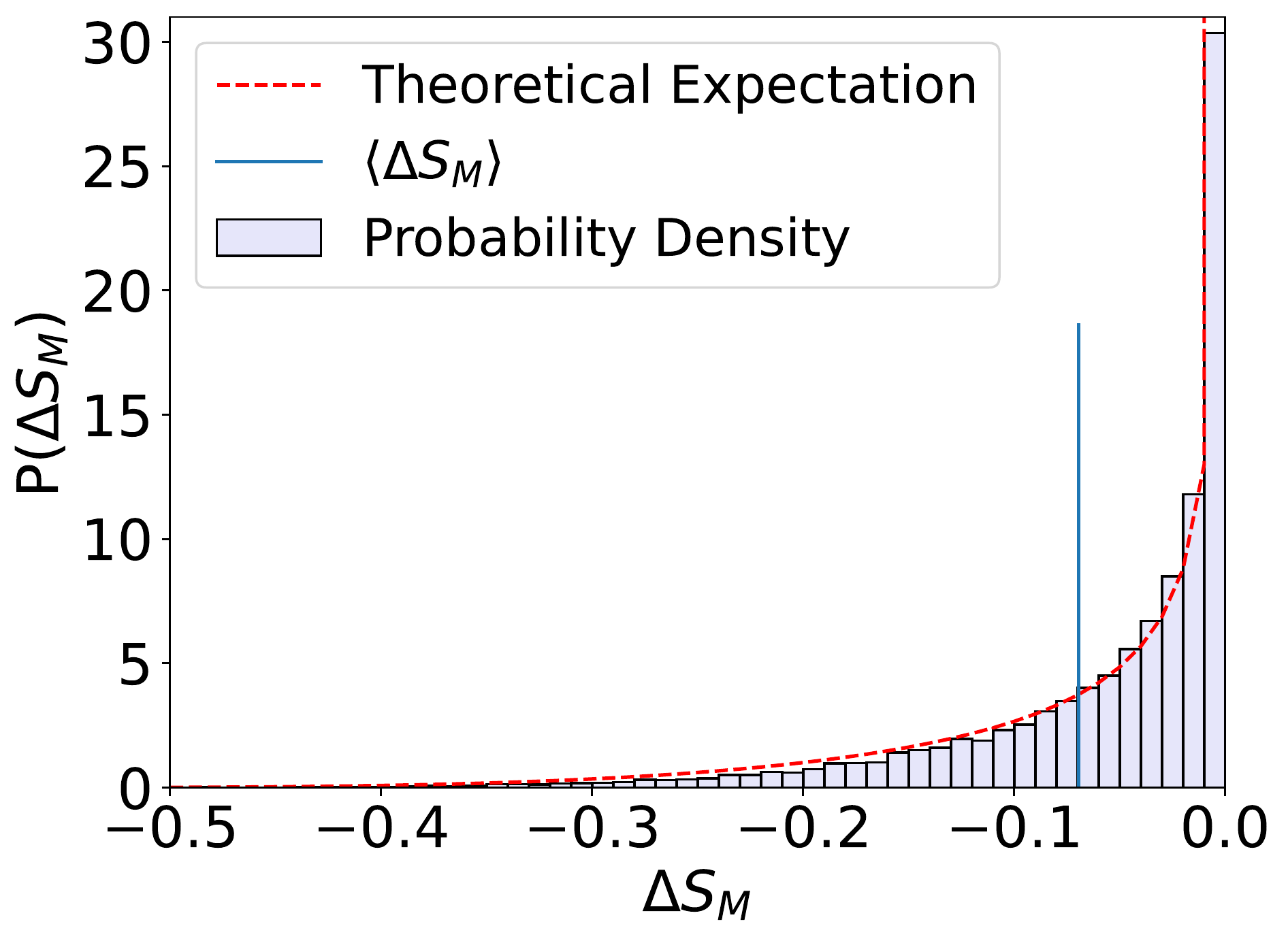}
  \includegraphics[scale=0.45]{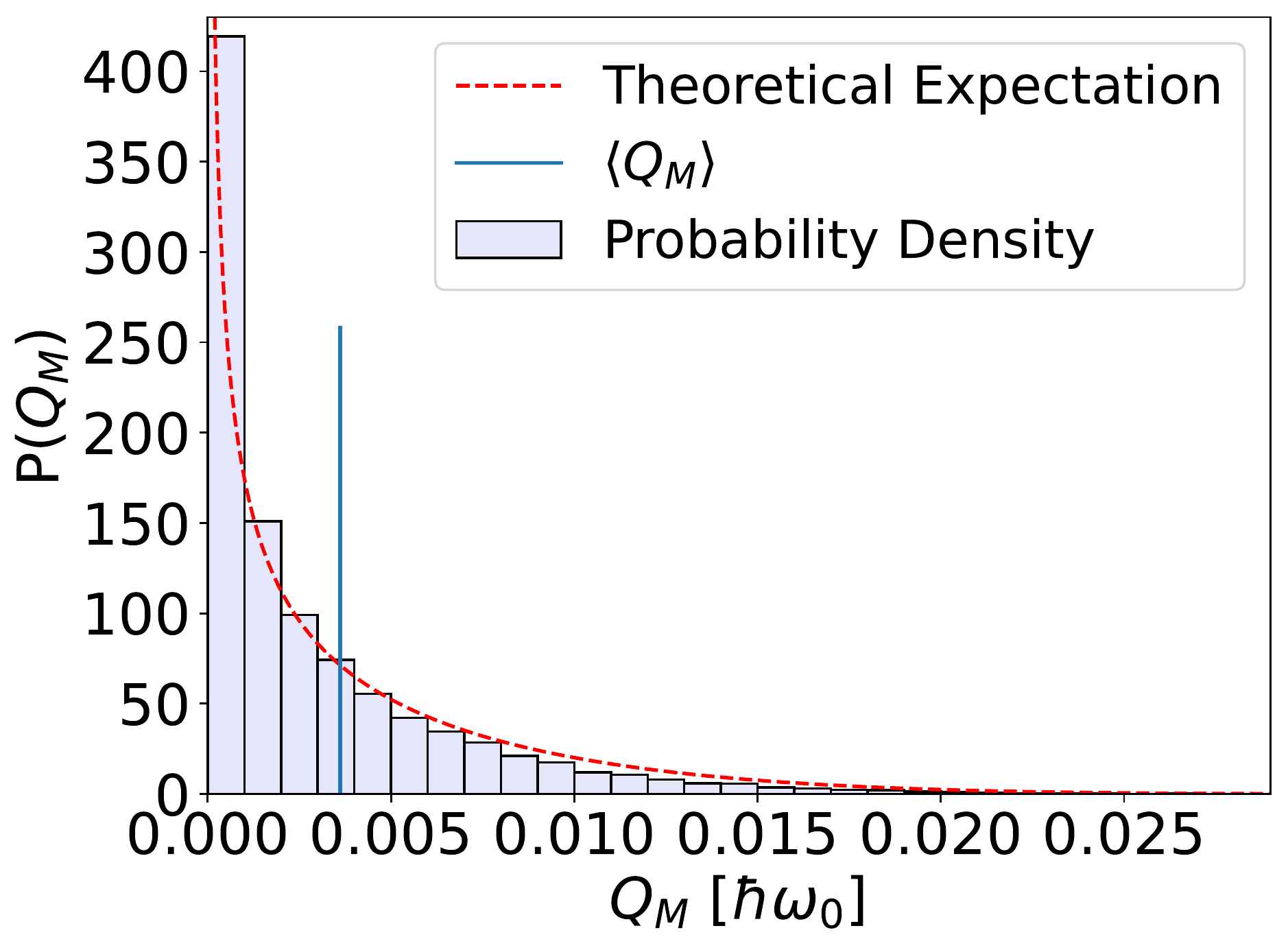}
 \caption{Probability distribution of the work extraction (top panel), entropy production between the final measurement and initial state of the qubit (middle panel), and the change in energy due to a series of measurements (bottom panel) for $dt/\tau=0.01$ and $z_0 = -0.1$. The simulation is done for 15 sequential continuous measurements with feedback application only at the end. The distributions are for 20,000 simulations. We take, $\hbar\omega_0=k_{\rm B}T$.}
 \label{fig:HistW_weak}
\end{figure}
To do so, we make use of the knowledge of probability density of $Q$, which can be expressed as \cite{dressel2017arrow,jayaseelan2021quantum}

\begin{equation}
P(Q)=\sqrt{\frac{\tau}{2\pi \cal T}}\frac{e^Q}{\sqrt{e^Q-1}}\,e^{\Big(-\frac{\cal T}{2\tau}-\frac{\tau}{2\cal T}\left[\cosh^{-1}(e^{Q/2})\right]^2\Big)},
\label{eq:dis_Q}
\end{equation}
where $\cal T$$=n dt$ such that $n$ is the number of independent measurements made in one simulation and $dt$ is the time interval between two sequential measurements. 
 These finite-time distributions have been studied both experimentally and theoretically in both superconducting qubits as well as cold atoms~\cite{dressel2017arrow,jayaseelan2021quantum,harrington2019characterizing}. To derive the finite-time statistics of other thermodynamic variables, we may make use of the (corresponding time-continuous limit of) identities we derived in Sec.~\ref{sec2}. For example, the probability distribution of extractable work in arbitrary finite-time can be derived from the probability distribution of the exponential of the quantum measurement arrow of time. This is given by (see Appendix~\ref{app:A} for details),
\begin{equation}
P(W_{\rm ext})=-\frac{4e^{Q}}{\hbar\omega_0}\frac{1}{z_0e^{Q/2}+\frac{z_0^2-1}{\sqrt{1+\left(z_0^2-1\right)e^{-Q}}}}P(Q),
\end{equation}
where using Eq.~\ref{eq:wextQ} we obtain following relation between $e^{-Q}$ and $W_{\rm ext}$
\begin{equation}
e^{-Q}=\left[\frac{2W_{\rm ext}z_0}{\hbar\omega_0}+\sqrt{1+\frac{4W_{\rm ext}^2}{\hbar^2\omega_0^2}(z_0^2-1)}\right]^2.
\end{equation}
Similarly, the probability distribution for the measurement heat $Q_{M}$ can be expressed as
\begin{equation}
P(Q_{M})=-\frac{4e^{Q/2}}{\hbar \omega_0 z_0}P(Q).
\end{equation}
From Eq.~(\ref{eq:delEQ}), we have $e^{-Q}=\left(2Q_{M}/\hbar \omega_0z_0+1\right)^2$. The average heat generated by the measurement can be expressed as
\begin{equation}
\langle Q_{M}\rangle=\frac{1}{2}\hbar\omega_{0}z_0\big(e^{-\frac{\delta t}{2\tau}}-1\big).
\label{eq:av_meas_en}
\end{equation} We can use the same procedure to derive the theoretical expectation for the probability distributions of the change in entropy as well.

In Fig.~\ref{fig:HistW_weak}, we compare the probability distribution plots and the theoretical expectation for the probability distributions for the work extraction (top panel), the change in entropy after the final measurement (middle panel) and the total energy provided by the measurement (bottom panel) for 20,000 simulations of the work extraction process. We show that, for weak continuous measurements, the engine is more likely to extract work near zero and its probability to extract higher work decreases as we approach the work extraction for strong measurement limit ($\frac{1}{2}\hbar\omega_0$). The entropy of the qubit decreases after all the measurements. Hence, $\Delta S_M$ is negative. For weak measurements, we are most likely to get no change in entropy and the probability of change in entropy decreases as the entropy decreases further. In the case of energy supplied by measurement, similar to the case of work extraction and entropy change we are most likely to find $Q_{M}$ near $0$, with its average given by Eq.~(\ref{eq:av_meas_en}). We also show that our theoretical expectations accurately match the simulations.

\section{Conclusions}\label{conc}
We investigated the thermodynamic as well heat exchange properties of a single qubit based device driven by weak quantum measurements. We find interesting statistical connections between the relevant thermodynamic variables, work, heat, entropy production, and the demon's perceived arrow of time. Considering time-continuous weak quantum measurements, we derive the exact finite-time statistics of work, heat and entropy changes, and relate them to the known statistics of the quantum measurement arrow of time.

Our work has implications for both understanding the fundamental links between work, heat, entropy, and information flows in simple quantum devices, the constraints imposed on them by the principles of thermodynamics,  as well as the potential to probe them in feasible experiments. Both superconducting quantum circuits and ultra-cold atoms serve as immediate platforms where the above discussed identities can be probed in experiments.  The results discussed here also opens new directions of research towards achieving on-demand thermal control in simple quantum systems, for example, by controlling the accessible information flows (by measurements and feedback operations) across a chain of qubits in such a way that they determine the heat and entropy currents below a certain threshold. We defer this analysis to a future work. 

\begin{acknowledgments}
This work was supported by the U.S. Department of Energy (DOE), Office of Science, Basic Energy Sciences (BES), under Award No. DE-SC0017890. The work of SKM was supported in part by the Wallenberg Initiative on Networks and Quantum Information (WINQ). Nordita is partially supported by Nordforsk.
\end{acknowledgments}
\ \\

\appendix
  
\section{Linear entropy production and the distribution of work extraction}
\label{app:A}
The linear entropy can be defined as
\begin{equation}
S_L(\rho)=2\left(1-{\rm Tr}\left\{\rho\right\}^2\right).
\end{equation}
If $\rho_{i}^{\rm th}$ and $\rho_{M\pm}$ are the initial thermal state and the state of the qubit after measurement respectively, the linear entropy and the quantum measurement arrow of time satisfy the following relation
\begin{equation}
S_L(\rho_{M\pm})=\exp(-Q)S_L(\rho_i^{\rm th}).
\label{eq:lin_rel_ent}
\end{equation}
Averaging over many realizations, we obtain following equality
\begin{equation}
\left\langle e^{-Q+\Delta F}\right\rangle=1,
\end{equation}
where $\Delta F=\log S_L(\rho_i^{\rm th})-\log S_{L}(\rho_{M\pm})$ gives the logarithmic difference between the linear entropies of the initial state and the state after the measurement. In the spirit of Refs.~\onlinecite{manikandan2019,harrington2019characterizing,jayaseelan2021quantum}, the above result can be understood as a new ``fluctuation theorem" relating the arrow of time to linear entropy changes in the measurement process, when the initial states are strictly impure.

For the continuous weak measurement case, Eq.~(\ref{eq:lin_rel_ent}) can be rewritten as
\begin{equation}
\tilde{S}_L=\frac{S_L(\rho_{M\pm})}{S_L(\rho_i^{\rm th})}=e^{-Q}=\sech^2\left(\frac{\delta t r}{\tau}\right).
\end{equation}
Using Eq.~\ref{eq:dis_Q}, the probability density of $\tilde{S}_L$ can be expressed as
\begin{equation}
P(\tilde{S}_L)=-e^{Q}P(Q).
\end{equation}
Since the work extraction can be written in terms of $\tilde{S}_L$ as
\begin{equation}
W_{\rm ext}=\frac{\hbar \omega_0}{2} \left(z_0\sqrt{\tilde{S}_L}+\sqrt{1+\left(z_0^2-1\right)\tilde{S}_L}\right).
\end{equation}
The distribution for work can be written as
\begin{equation}
P(W)=\frac{4}{\hbar\omega_0}\frac{1}{\frac{z_0}{\sqrt{\tilde{S}_L}}+\frac{z_0^2-1}{\sqrt{1+\left(z_0^2-1\right)\tilde{S}_L}}}P(\tilde{S}_L).
\end{equation}

\bibliography{paper}

\end{document}